# Gas phase synthesis of non-bundled, small diameter single-walled carbon nanotubes with near-armchair chiralities

K. Mustonen[1], P. Laiho[1], A. Kaskela[1], Z. Zhu[1], O. Reynaud[1], N. Houbenov[1], Y. Tian[1], T. Susi[2], H. Jiang[1], A.G. Nasibulin[1,3], E.I. Kauppinen[1,a)]

[1] *Department of Applied Physics, Aalto University School of Science, P.O. Box 15100, FI-00076 Aalto, Finland*

[2] *University of Vienna, Faculty of Physics, Boltzmanngasse 5, A-1090 Vienna, Austria*

[3] *Skolkovo Institute of Science and Technology, 100 Novaya st., Skolkovo, Odintsovsky district, Moscow Region, 143025, Russia*

a) Electronic mail: esko.kauppinen@aalto.fi

We present a novel floating catalyst synthesis route for individual, i.e. non-bundled, small diameter single-walled carbon nanotubes (SWCNTs) with a narrow chiral angle distribution peaking at high chiralities near the armchair species. An *ex situ* spark discharge generator was used to form iron particles with geometric number mean diameters of 3-4 nm and fed into a laminar flow chemical vapour deposition reactor for the continuous synthesis of long and high-quality SWCNTs from ambient pressure carbon monoxide. The intensity ratio of G/D peaks in Raman spectra up to 48 and mean tube lengths up to 4 μm were observed. The chiral distributions, as directly determined by electron diffraction in the transmission electron microscope, clustered around the (n,m) indices (7,6), (8,6), (8,7) and (9,6), with up to 70% of tubes having chiral angles over 20°. The mean diameter of SWCNTs was reduced from 1.10 to 1.04 nm by decreasing the growth temperature from 880 to 750 °C, which simultaneously increased the fraction of semiconducting tubes from 67 to 80%. Limiting the nanotube gas phase number concentration to ~$10^5$ $cm^{-3}$ successfully prevented nanotube bundle formation that is due to collisions induced by Brownian diffusion. Up to 80 % of 500 as-deposited tubes observed by atomic force and transmission electron microscopy were individual. Transparent conducting films deposited from these SWCNTs exhibited record low sheet resistances of 63 Ω/□ at 90 % transparency for 550 nm light.

Many electronic applications of single-walled carbon nanotubes (SWCNT), including field effect transistors (FETs) – both single-tube[1] and percolating[2] – and nano-electromechanical systems,[3] benefit from clean, high-quality and non-aggregated SWCNTs. Although catalysts are mostly immobile in substrate-supported chemical vapor deposition (CVD) techniques, enabling the growth of well-separated tubes,[4] usable support substrates are mainly limited to oxides that are able to withstand elevated temperatures required by the SWCNT growth. This prevents *in situ* growth on temperature sensitive substrates required e.g. by flexible and transparent conductive films (TCFs), necessitating additional dispersion and deposition steps that often lead to a degradation of the tubes' intrinsic properties due to the use of chemical surfactants and tube cutting during sonication.[5]

For such applications and for many fundamental studies, floating catalyst chemical vapor deposition (FC-CVD) methods such as high pressure CO disproportionation[6] (HiPco) and ferrocene vapor decomposition[7] offer distinct advantages, for they provide a possibility to directly deposit high quality SWCNTs onto any substrate.[8] However, apart from using post-synthesis electrostatic filtration to remove electrically charged bundles from the aerosol flow,[9] these methods have so far failed to overcome the bundle formation issue, which has limited their usefulness.

We attribute this drawback to the lack of efficient means for controlling the gas phase number concentration (N) of SWCNTs during growth, which leads to mutual collisions and bundle formation due to Brownian diffusion. This collision rate of aerosol particles is related to mobility diameter ($D_M$), ambient temperature (T) and carrier gas viscosity ($\eta$) through[10]

$$\frac{dN}{dt} = -K(D_M, T, \eta)N^2, \qquad (1)$$

where K is the coagulation coefficient. Since in a particular growth condition, the dimensions of the nanotubes and the temperature do not vary, the collision rate is simply proportional to $N^2$. Unfortunately, the mobility diameters of SWCNTs are not directly known, although e.g. Kim et al. have estimated that multiwalled CNTs of 15 nm diameters and 100-700 nm lengths have $D_M$ of approximately 50-120 nm.[11] For SWCNTs with similar lengths these are likely crude overestimates, as for non-isometric fibrous structures the diameter is an important parameter defining the mobility.[12] Thus, we estimate the $D_M$ for SWCNTs of around 10-30 nm, resulting in coagulation coefficients between $2.6 \times 10^{-9}$ and $9.0 \times 10^{-9}$ cm$^3$ s$^{-1}$, therefore implying that in the reactor the number concentration should be kept well below $10^6$ cm$^{-3}$ to limit bundle formation during a typical 10 s gas flow residence time.

The floating catalyst CVD technique we present here was designed to control $N_{CNT}$ by decoupling the catalyst and nanotube growth steps into two subsequent processes. The catalyst particles are formed by a spark discharge generator[13] (Figure 1), in which Fe (purity 99.95%, Goodfellow, UK) evaporates from a pair of rod-shaped electrodes (spacing L=0.5-1.0 mm, diameter 2-5 mm), constantly flushed by a high velocity $N_2$ jet (nozzle diameter 2.5 mm, velocity 200-1500 ms$^{-1}$) held 4 mm from the gap. The spark discharge gap capacitance (C=0.47-47 nF) is recharged using a high-voltage source (20 kV, Matsushita, Japan) through a ballast resistor (R=1.5 MΩ) until a discharge occurs and the process starts over. Hence, the recharge voltage (typically 2-7 kV) defines the recharge current and the discharge frequency (f, typically 0.1-1.0 kHz). The discharges between the electrodes generate plasma that evaporates the catalyst material. The electrode mass loss rate is proportional to the frequency of constant energy discharges, whereas the resulting mass concentration of Fe per unit volume of the $N_2$ dilution gas is proportional to the volumetric flow rate (typically Q=6-

$45\times10^3$ cm$^3$min$^{-1}$). Each discharge creates an approximately constant number ($n_0$) of primary particles, or condensation nuclei, and thus the number concentration of primary particles $N_p$ becomes[13]

$$N_p = n_0 \frac{f}{Q}. \qquad (2)$$

According to Schwyn *et al.*, the geometric mean diameter ($D_g$) of primary particles from such discharges is on the order of 1-2 nm, while $n_0$ is between $10^5$ and $10^6$ with a capacitance of 2.2 nF and a gap of 1 mm.[13] However, regardless of the high volumetric dilution flow, the primary particles aggregate due to their extremely high Brownian diffusivity, resulting in a lognormal diameter distribution with a geometric mean diameter larger than 3 nm.

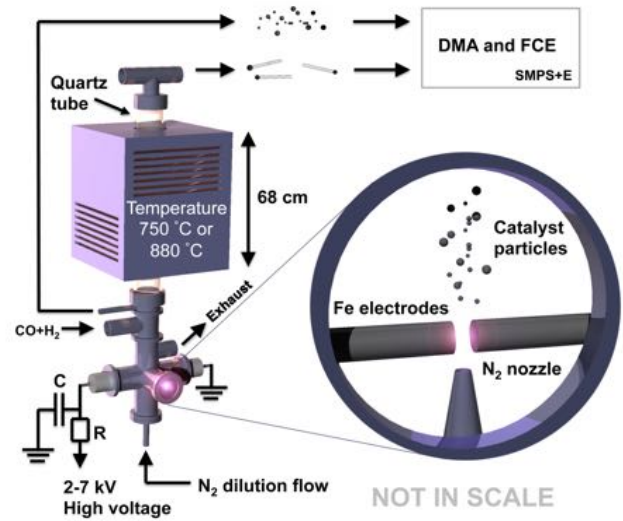

Fig.1. A schema of the synthesis reactor. The spark generator consists of a pair of iron electrodes separated by a discharge gap, continuously flushed by a high-velocity N$_2$ jet. The discharges evaporate metal from the electrodes, forming catalyst particles that are subsequently fed into the vertical CVD reactor consisting of a quartz tube in a high temperature furnace. An SMPS+E (scanning mobility particle sizer with electrometer) aerosol size classifier consisting of a differential mobility analyser (DMA) and Faraday cup

electrometer (FCE) is used to determine the catalyst number concentrations (N) and geometric mean diameter ($D_g$) prior to introduction into the reactor, and those of the SWCNTs at the reactor outlet.

Prior to introducing the catalyst particles into the CVD reactor, their number size distribution (NSD) is measured using an SMPS+E system (scanning mobility particle sizer with electrometer, GRIMM Aerosol Technic GmbH, Germany) consisting of a differential mobility analyzer (Vienna-type Nano-DMA, length 15 mm) and Faraday cup electrometer (FCE, sensitivity 0.1 fA). Figure 2a shows typical catalyst NSDs at recharge voltages of 2.2, 2.8 and 3.2 kV. Practical experience operating the system has shown us that by adjusting Q, L, f, C and recharge voltages, the geometric mean diameter $D_g$ can be tuned from 3 nm upwards. The catalyst collisions and the following aggregation, however, also set the practical upper limit of the CNT number concentration to ~$10^6$ cm$^{-3}$; further increasing the concentration results in particle aggregate formation on the expense of nanotube yield. Figure 2a illustrates the tendency of particle growth together with the increasing number concentration. Thus, in order to both avoid significant deposits of inactive catalyst aggregates on the nanotubes and to prevent their bundling, the desired catalyst $D_g$ is between 3-4 nm and N between $10^5$ to $10^6$ cm$^{-3}$.

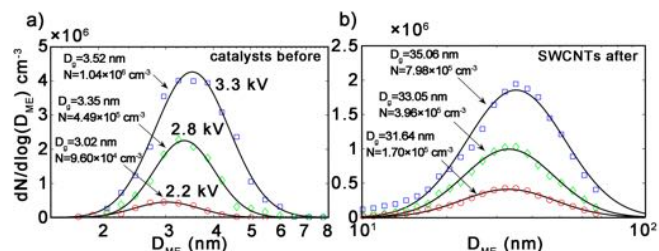

Fig. 2. Catalyst particle and SWCNT number size distributions (NSDs) measured with the DMA with different spark generator settings. **a)** Catalyst NSDs acquired with spark settings

C=47 nF, L=0.5 mm, Q=45 lpm, and $U_R$=2.2, 2.8 and 3.3 kV, having geometric mean diameters ($D_g$) of 3.02, 3.35 and 3.52 nm, and corresponding number concentrations (N) of $9.60×10^4$, $4.49×10^5$ and $1.04×10^6$ cm$^{-3}$, respectively. **b)** The corresponding SWCNT NSDs, demonstrating a correlation between the catalyst and SWCNT number concentrations. $D_{ME}$ on the logarithmic x-axis is the electrical mobility diameter.

Downstream the spark generator, the particles carried by the flowing $N_2$ (limited to 200 cm$^{-3}$min$^{-1}$, the rest going to exhaust) are mixed with carbon monoxide (CO, 250 cm$^{-3}$min$^{-1}$) and hydrogen ($H_2$, 50 cm$^{-3}$min$^{-1}$) and introduced into a laminar flow (Reynolds number 25) vertical CVD reactor consisting of a quartz tube (inner diameter 22 mm) placed inside a high-temperature furnace. The FC-CVD reactor flows were designed using computation fluid dynamics (CFD) calculations to prevent turbulence, as that may significantly increase the particle collision rate[10] (the CFD solutions are presented in Ref. 14). For SWCNT growth, the reactor temperature is set either to 750 or 880 °C, where carbon is exclusively released through catalytic CO disproportionation and hydrogenation reactions on the surfaces of the catalysts, resulting in the growth of clean nanotubes.[7,15] Crucially, the CNT number concentration is directly proportional to the concentration of catalyst in the active size range (Figure 2a and b) and can therefore be defined prior to synthesis.

Scanning electron microscopy (SEM, Zeiss Sigma VP) was used to deduce the SWCNT lengths from tubes thermophoretically[16] deposited onto p-doped Si/SiO$_2$ wafers. The lengths exhibited the expected lognormal distributions[8] with geometric means of 1.95 μm for the growth temperature of 750 °C and 4.03 μm for 880 °C, indicating an approximately two-fold difference in growth rates since the residence time in the reactor varies only slightly. The length distributions can be found in Ref. 14.

A high-resolution transmission electron microscope (TEM, 2×$C_s$-corrected JEOL JEM-2200FS, JEOL Ltd. Japan) operated at 80 kV was used to measure the tube diameters and bundle sizes, and chiral distributions determined via electron diffraction. For TEM characterization, the CNTs were directly deposited from the aerosol flow onto TEM grids using a collection time of 3 minutes with N set to ~2×10$^5$ cm$^{-3}$ to minimize gas phase collisions. According to TEM observations, the as-grown SWCNTs were clean and had relatively small mean diameters varying with growth temperature from 1.04±0.19 nm at 750 °C to 1.10±0.26 nm at 880 °C, as shown in Figure 3a-d and f. The larger diameter at 880 °C may be attributed to enhanced CO disproportionation, releasing more carbon to saturate slightly larger diameter catalyst particles which therefore become active for SWCNT growth. The mean diameters of the nanotubes were similar to CoMoCat[17] and to HiPco[6] at ~ 0.8 nm and ~1.0 nm, respectively. Figure 3 g-h and Table I show how the (n,m) distributions were clustered at high chiral angles around the (7,6) and (8,6) species at a growth temperature of 750 °C, and near (8,7) and (9,6) at 880 °C; up to 67% of our tubes had chiral angles ≥20°, which is a significantly narrower angle distribution than that produced by the HiPco method.[18] In fact, our chiral angle distributions were even narrower than those of CoMoCat SWCNTs,[19,20] typically considered to have the narrowest distribution of all standard CVD methods. The observed fraction of semiconducting tubes in the 880 °C sample was the expected 67%, whereas the 750 °C had a larger fraction of 80%.

Table I. SWCNT chiral angle populations measured by electron diffraction (~90 tube statistics for each temperature for spark, 265 tubes for CoMoCat).[19]

|  | 10° ≥ θ | 20° ≥ θ > 10° | θ > 20° |
| --- | --- | --- | --- |
| Spark 750 °C | 9 (12%) | 19 (21%) | 59 (67%) |
| Spark 880 °C | 8 (8.5%) | 30 (31.1%) | 58 (60.4%) |
| NIST CoMoCat | 39 (15%) | 82 (30%) | 148 (55%) |

Importantly, careful and systematic observations of the bundles deposited onto TEM grids revealed that 60% of the tubes were individual, strikingly different from other FC-CVD methods reported in the literature.[6,7,15] Moreover, while some small bundles were also present in the sample, only about 10% contained more than two individual tubes, as illustrated in Figure 3e.

The TEM studies were supplemented by atomic force microscopy (AFM, Dimension 5000, Veeco Instruments Inc., USA) of SWCNTs thermophoretically[16] deposited onto mica substrates and imaged using settings optimized for nanotubes.[21] In general, similar to TEM, the bundles and individual tubes were sparsely deposited and clearly resolved from the substrate, as is evident from Figure 4. The heights of the bundles are best illustrated via profiler tools (profiles 1-6 shown in Figure 4 and the rest in Ref. 14) drawn in the scan window. The representative sample of bundles in Figure 4 exhibits heights between 1 and 1.3 nm, whereas a larger statistical sample (n≈400) revealed a mean height of 1.21 nm, with up to 80% of the bundles having heights <1.4 nm. Importantly, the bundle heights were almost identically distributed to the TEM data, as is evident by comparing the histograms in Figures 4 and 3e.

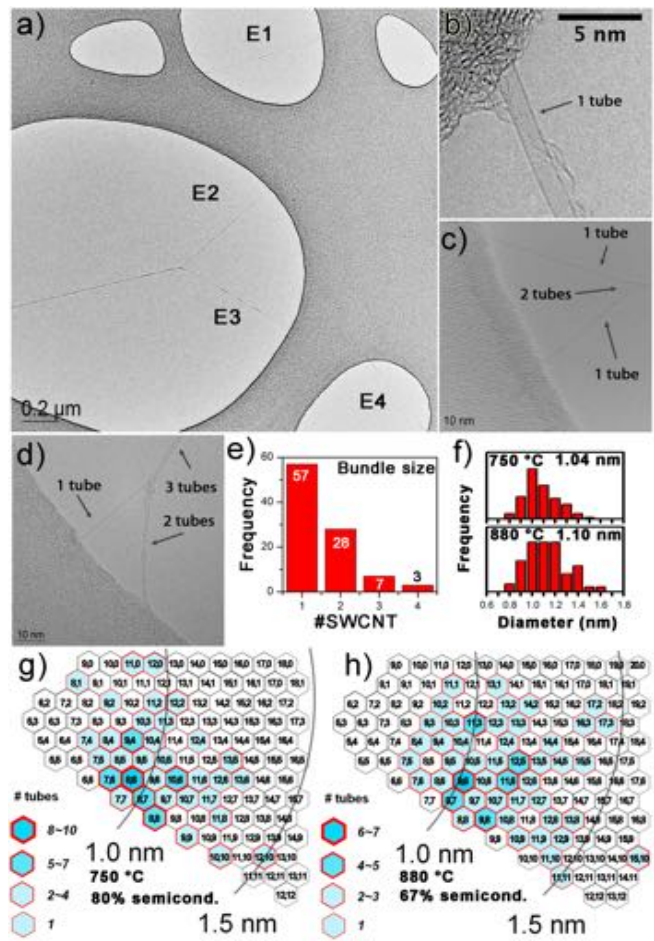

Fig. 3. TEM characterization of as-deposited SWCNTs. **(a)** An example of the typical SWCNT morphology with a 3 minute collection time, exhibiting a high fraction of individual tubes. Electron diffraction patterns of the individual tubes E1-E4 are shown in Ref. 14. **(b-d)** Micrographs showing typical individual tubes and small bundles (2-3 tubes; synthesized at 880 °C), with very little surface contamination and **(e)** bundle size statistics showing 60% of individual tubes. (**f**) Diameter statistics of the tubes synthesized both at 750 and 880 °C, and **(g-h)** the chiral angle maps.

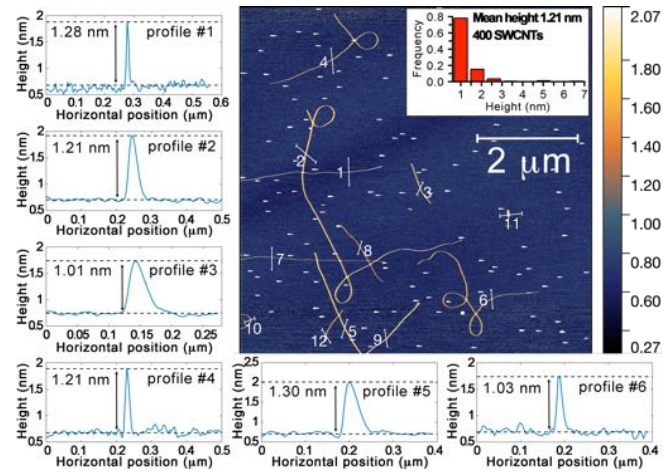

Fig. 4. AFM characterization of SWCNTs synthesized at 880 °C with $N \approx 10^5$ cm$^{-3}$. The scan window size is 7×7 μm². The vertical height profiles of the cross sections 1-6 are shown around the scan window, representing heights between 1 and 1.3 nm (profiles 7-12 can be found in Ref. 14). The histogram shows the statistics of 400 height profiles, with the mean at 1.21 nm.

Optical absorption spectroscopy (Lambda 950 spectrometer, PerkinElmer Inc., USA) was used to further characterize the properties of SWCNTs. For optical measurements, the SWCNTs were collected on nitrocellulose membrane filters (Millipore, France) and press-transferred[8] onto 1 mm thick quartz windows. The background-subtracted optical absorption spectra are shown in Figure 5a. Fitting[22] the spectra yielded geometric mean diameters of 1.05 nm and 1.13 nm for the SWCNTs grown at 750 °C and 880 °C respectively, in excellent agreement with the TEM measurements (Figure 3f). The fitted diameter distributions can be found in Ref. 14. The very well resolved optical transition peaks in the spectra indicate a high fraction of individual SWCNTs and relatively narrow chirality distributions. The most prominent absorption peaks (Figure 5a) show that the (7,6), (8,6) and (8,7) species dominate in 750 °C sample, whereas the (8,7), (9,7) and (9,8) chiralities are strongly represented in the 880 °C sample. Resonant Raman measurements (Labram HR spectroscope, Horiba Ltd. Japan) with a laser wavelength of 633 nm further verified that the tubes were of very high quality. This was indicated by high G/D intensity ratios[23] between ~42 and ~48, shown in Figure 5c-d, along with intense radial breathing modes (RBMs). Due to the reverse relationship of RBM frequency and diameter, a shift of Raman RBM intensities towards higher frequency in the lower temperature sample indicates smaller tube mean diameters, again in good agreement with TEM and optical absorption measurements.[24]

Finally, to demonstrate the application potential of the spark technique, TCFs out of 4 μm long individual tubes were fabricated using press-transfer and treated with a strong solution of nitric acid ($HNO_3$). Figure 5b shows a performance of 63 Ω/□ at 90% transparency for 550 nm light at ambient conditions, being among the best values reported for CNT TCFs.[25] The substrate contribution was subtracted by placing a clean quartz window to the reference beam path, and the sheet resistances evaluated with a Jandel Ltd. 4-point probe at a 60 gram needle loading. We note that while films of 90% transparency with a diameter of a few centimeters take several hours to deposit with the current small-scale reactor, prospects for scaling up the setup are quite straightforward.

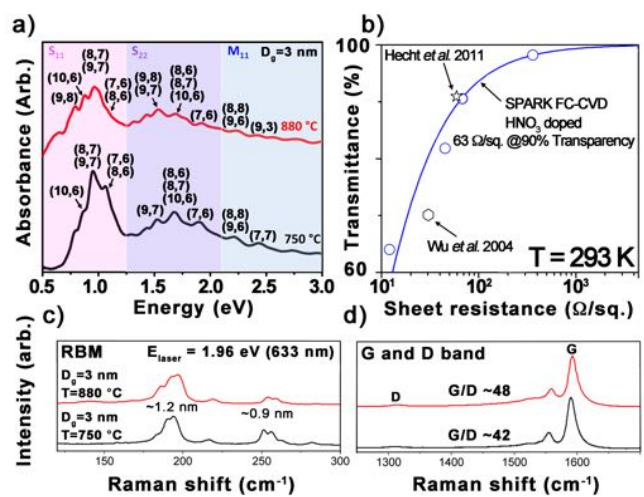

Fig. 5. Optical properties of the synthesized SWCNTs. **(a)** Optical absorption spectra of the SWCNTs synthesized at 750 and 880 °C, showing prominent and sharp resonant transition peaks dominated by (7,6), (8,6) and (8,7) chiralities in the 750 °C sample and (8,7), (9,7) and (9,8) in the 880 °C sample. $S_{11}$, $S_{22}$ and $M_{11}$ represent the energy ranges of the lowest-energy semiconducting, second-lowest energy semiconducting, and lowest-energy metallic absorption peaks, respectively. **(b)** The electro-optical performance of $HNO_3$-treated spark FC-CVD tubes with 63 Ω/□ sheet resistance at a 90% transparency measured at ambient conditions and at the 550 nm wavelength. **(d)** RBMs and **(e)** G and D bands of the SWCNTs grown at 750 °C and 880 °C excited by 633 nm laser.

To summarize, we have described a unique floating catalyst system for the synthesis of mostly individual, long and small diameter SWCNTs with narrow chiral angle distributions. The chiralities were clustered around (7,6) and (8,6) at a synthesis temperature of 750 °C and (8,7) and (9,6) at 880 °C, with up to 70% of tubes having chiral angle between 20° and 30°. The high SWCNT individuality was achieved after Brownian diffusion was identified as the primary cause for CNT bundle formation in the gas phase and prevented by limiting the CNT number concentration to ~$10^5$ cm$^{-3}$. Therefore, especially in the floating catalyst routes, the controllability of the catalyst number concentration is essential, dictating the morphology of the end product all the way from individual tubes to large bundles. A high level of individual nanotubes can only be achieved by accepting a trade-off with the rate of CNT production. We have shown that at reactor number concentrations around $10^5$ cm$^{-3}$, a fraction of up to 80% of individual SWCNTs can be achieved on surfaces – a figure that is likely reduced from the gas phase proportion by collisions during deposition. The production rate remains acceptable, though certainly depending on the choice of application; a greater yield is necessary for bulk production. Nevertheless, even with the current capacity, we successfully demonstrated the fabrication of transparent conductive films with a record high performance of 63 Ω/□ at 90 % transparency, highlighting the application potential of our spark FC-CVD system. Besides transparent conductors, SWCNTs for many specific applications such as field effect transistors, nano-electromechanical systems, and quantum oscillators can readily be realized with the current system even at concentrations well below $10^5$ cm$^{-3}$, potentially producing close to 100% as-deposited individual tubes.

**Acknowledgements**

The research leading to these results has received funding from the European Union Seventh Framework Programme (*FP7/2007-2013*) under *grant agreement* n° 604472 (IRENA project) and n° 314068 (TREASORES project), by the Aalto Energy Efficiency (AEF) program through the MOPPI project, and from TEKES projects CARLA and USG and Academy of Finland (HISCON n° 276160). A.G.N. was partially supported by the Ministry of Education and Science of the Russian Federation (Project DOI: RFMEFI58114X0006), and T.S. by the Austrian Science Fund (FWF) through grant M 1497-N19, by the Finnish Cultural Foundation, and by the Walter Ahlström Foundation. This work made use of the Aalto University Nanomicroscopy Center (Aalto-NMC) premises. The personnel of the National Nanomicroscopy Center of Aalto University are gratefully acknowledged for useful discussions and assistance.